\documentclass[aps,pra,10pt,showpacs,eqsecnum,amsmath,twocolumn,floatfix,superscriptaddress]{revtex4-1}
\usepackage[dvipsnames]{xcolor}
\usepackage[colorlinks=true,linkcolor=NavyBlue,urlcolor=NavyBlue,citecolor=NavyBlue,breaklinks]{hyperref}
\usepackage{amsmath}
\usepackage{amsfonts}
\usepackage{graphicx}
\newcommand{\bs}[1]{\boldsymbol{#1}}

\newcommand{\bk}[1]{\left(#1\right)}
\newcommand{\Bk}[1]{\left[#1\right]}
\newcommand{\BK}[1]{\left\{#1\right\}}

\newcommand{\expect}{\operatorname{\mathbb E}}
\newcommand{\erfc}{\operatorname{erfc}}
\begin{document}

\title{Optimal signal processing for continuous qubit readout}

\author{Shilin Ng}
\affiliation{Department of Physics, National University of Singapore,
  2 Science Drive 3, Singapore 117551}

\author{Mankei Tsang}
\email{eletmk@nus.edu.sg}
\affiliation{Department of Electrical and Computer Engineering,
  National University of Singapore, 4 Engineering Drive 3, Singapore
  117583}

\affiliation{Department of Physics, National University of Singapore,
  2 Science Drive 3, Singapore 117551}
\date{\today}


\begin{abstract}
  The measurement of a quantum two-level system, or a qubit in modern
  terminology, often involves an electromagnetic field that interacts
  with the qubit, before the field is measured continuously and the
  qubit state is inferred from the noisy field measurement.  During
  the measurement, the qubit may undergo spontaneous transitions,
  further obscuring the initial qubit state from the
  observer. Taking advantage of some well known techniques in
  stochastic detection theory, here we propose a novel signal
  processing protocol that can infer the initial qubit state optimally
  from the measurement in the presence of noise and qubit
  dynamics. Assuming continuous quantum-nondemolition measurements
  with Gaussian or Poissonian noise and a classical Markov model for
  the qubit, we derive analytic solutions to the protocol in some
  special cases of interest using It\={o} calculus. Our method is
  applicable to multi-hypothesis testing for robust qubit readout and
  relevant to experiments on qubits in superconducting microwave
  circuits, trapped ions, nitrogen-vacancy centers in diamond,
  semiconductor quantum dots, or phosphorus donors in silicon.
\end{abstract}

\maketitle
\section{Introduction}
Consider a quantum two-level system, or a qubit in modern
terminology. According to von Neumann, measurement of a
qubit can be instantaneous and perfectly accurate, with two possible
outcomes and the qubit collapsing to a specific state depending on the
outcome \cite{wiseman_milburn}. In practice, this measurement model,
called a projective measurement, is an idealization. A qubit
measurement in real physical systems, such as superconducting
microwave circuits \cite{wallraff,lupascu,johnson12}, trapped ions
\cite{hume,myerson}, nitrogen-vacancy centers in diamond
\cite{neumann,robledo}, semiconductor quantum dots
\cite{elzerman,vamivakas}, and phosphorus donors in silicon
\cite{morello,pla}, is often performed by coupling the qubit to an
electromagnetic field, before the field is measured continuously. The
qubit state can only be inferred with some degree of uncertainty from
the noisy measurement. During the measurement, the qubit may also
undergo spontaneous transitions, which further obscure the initial
qubit state and complicate the inference procedure.  This qubit
readout problem is challenging but important for many
quantum information processing applications, such as quantum computing
\cite{nielsen}, magnetometry \cite{rondin}, and atomic clocks
\cite{schmidt05,chou}, which all require accurate measurements of
qubits. The choice of a signal processing method is crucial to the
readout performance. Refs.~\cite{gambetta07,danjou} in particular
contain detailed theoretical studies of qubit-readout signal
processing protocols.

In this paper, we propose a new signal-processing architecture for
optimal qubit readout by exploiting well known techniques in classical
detection theory
\cite{vantrees,kailath_poor,duncan_lr,kailath69,snyder}.  Following
prior work \cite{gambetta07,danjou}, we assume that the measurement is
quantum nondemolition (QND) \cite{braginsky,wiseman_milburn}, meaning
that a classical stochastic theory is sufficient
\cite{wiseman_milburn,qmfs,gough}. In addition to the Gaussian
observation noise assumed in Refs.~\cite{gambetta07,danjou}, we also
consider a Poissonian noise model \cite{snyder_miller}, which is more
suitable for photon-counting measurements
\cite{schmidt05,hume,myerson,neumann,robledo,vamivakas}. We find that
the likelihood ratio needed for optimal hypothesis testing can be
determined from the celebrated estimator-correlator formulas
\cite{kailath_poor,duncan_lr,kailath69,snyder,hypothesis}, which break
down the likelihood-ratio calculation into an estimator step and an
easy correlator step. The estimator turns out to have analytic
solutions in special cases of interest and simple numerical algorithms
in general.

Although our protocols and the ones proposed in
Refs.~\cite{gambetta07,danjou} should result in the same end results
for the likelihood ratio in the case of Gaussian noise, our analytic
solutions involve elementary mathematical operations and may be
implemented by low-latency electronics, such as analog or programmable
logic devices \cite{stockton02}, for fast feedback control and error
correction purposes \cite{wiseman_milburn}. This is in contrast to the
more complicated coupled stochastic differential equations recommended
by the prior studies.  Moreover, the prior studies never state whether
their stochastic equations should be interpreted in the It\={o} sense
or the Stratonovich sense, making it difficult for others to verify
and correctly implement their protocols. As the equations are
nonlinear with respect to the observation process, applying the wrong
stochastic calculus is likely to give wrong results
\cite{kailath_poor,gardiner,mao,snyder_miller}. Our work here, on the
other hand, makes explicit and consistent use of It\={o} calculus to
ensure its correctness. Our estimator-correlator protocol is also
inherently applicable to multi-hypothesis testing, which can be useful
for online parameter estimation and making the readout robust against
model uncertainties \cite{gambetta2001,chase,rbk10,wiebe,combes14}.

\section{\label{sec_hyp}Hypothesis testing}
Let $\{\mathcal H_m; m = 0,1,2,\dots,M-1\}$ be the hypotheses to be
tested. Given a noisy observation record $Z$, suppose that we use a
function $\tilde{\mathcal H}(Z)$ to decide on a hypothesis. Defining
the observation probability measure as $dP(Z|\mathcal H_m)$ and the
prior probability distribution as $P(\mathcal H_m)$, the average error
probability is
\begin{align}
P_e &\equiv \sum_m P(\mathcal H_m)
\int_{\tilde{\mathcal H}(Z)\neq\mathcal H_m}dP(Z|\mathcal H_m).
\label{Pe}
\end{align}
The decision rule that minimizes $P_e$ is to choose the hypothesis
that maximizes the posterior probability function
\cite{vantrees,berger}, which can be expressed as
\begin{align}
P(\mathcal H_m|Z) &= \frac{\Lambda(Z|\mathcal H_m)P(\mathcal H_m)}
{\sum_m\Lambda(Z|\mathcal H_m)P(\mathcal H_m)},
\label{bayes}
\end{align}
where we have defined
\begin{align}
\Lambda(Z|\mathcal H_m) &\equiv 
\frac{dP(Z|\mathcal H_m)}{dP(Z|\mathcal H_0)}
\label{lr}
\end{align}
as the likelihood ratio for $\mathcal H_m$ against $\mathcal H_0$, the
null hypothesis.  The minimum-error decision strategy thus boils down
to the computation of $\Lambda(Z|\mathcal H_m)$ for all hypotheses of
interest, and then finding the hypothesis that maximizes $P(\mathcal
H_m|Z)$, or equivalently
\begin{align}
\tilde{\mathcal H}(Z) &= 
\arg \max_{\mathcal H_m}\Bk{\ln\Lambda(Z|\mathcal H_m)+\ln P(\mathcal H_m)},
\end{align}
where $\ln\Lambda(Z|\mathcal H_m)$ is a log-likelihood ratio (LLR).
Many frequentist protocols also involve the computation of the LLR and
a likelihood-ratio test \cite{vantrees}.

\section{\label{gauss}Gaussian noise model}
\subsection{\label{process}Observation process}
\begin{figure}[htbp]
\centerline{\includegraphics[width=0.48\textwidth]{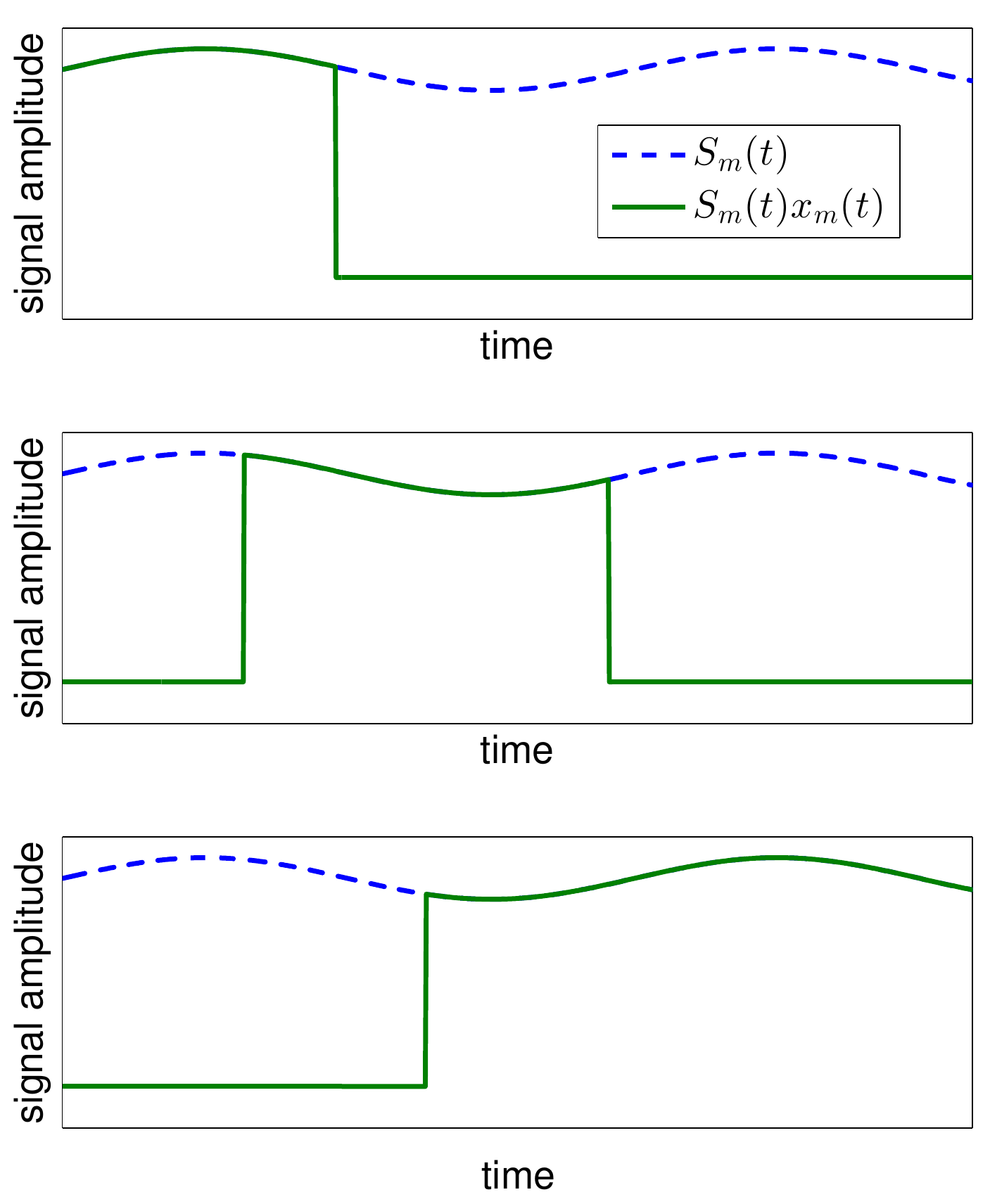}}
\caption{(Color online) Some example realizations of the signal
  component $S_m(t)x_m(t)$ of the observation process.  Given a
  hypothesis $\mathcal H_m$, $S_m(t)$ is a deterministic signal
  amplitude and $x_m(t)$ is a binary stochastic process. The
  axes are in arbitrary units.}
\label{trajectories}
\end{figure}

Assume that the observation process $z(t)$ conditioned on a
hypothesis is
\begin{align}
\mathcal H_m: \
z(t) &= S_m(t)x_m(t) + \xi(t),
\label{z}
\end{align}
where $S_m(t)$ is a deterministic signal amplitude assumed by the
hypothesis, $x_m(t)$ is a hidden stochastic process, $\xi(t)$ is a
zero-mean white Gaussian noise with covariance
\begin{align}
\expect\Bk{\xi(t)\xi(t')} &= R(t)\delta(t-t'),
\label{awgn}
\end{align}
$\expect$ denotes expectation, and $R(t)$ is the noise power, assumed
here to be the same for all hypotheses. It is possible to test other
values of noise power by prescaling the observation and redefining
$S_m(t)$. For qubit readout, the hypothesis should determine $S_m(t)$
and the statistics of $x_m(t)$; Fig.~\ref{trajectories} sketches a few
example realizations of the signal component $S_m(t)x_m(t)$.

\begin{figure}[htbp]
\centerline{\includegraphics[width=0.48\textwidth]{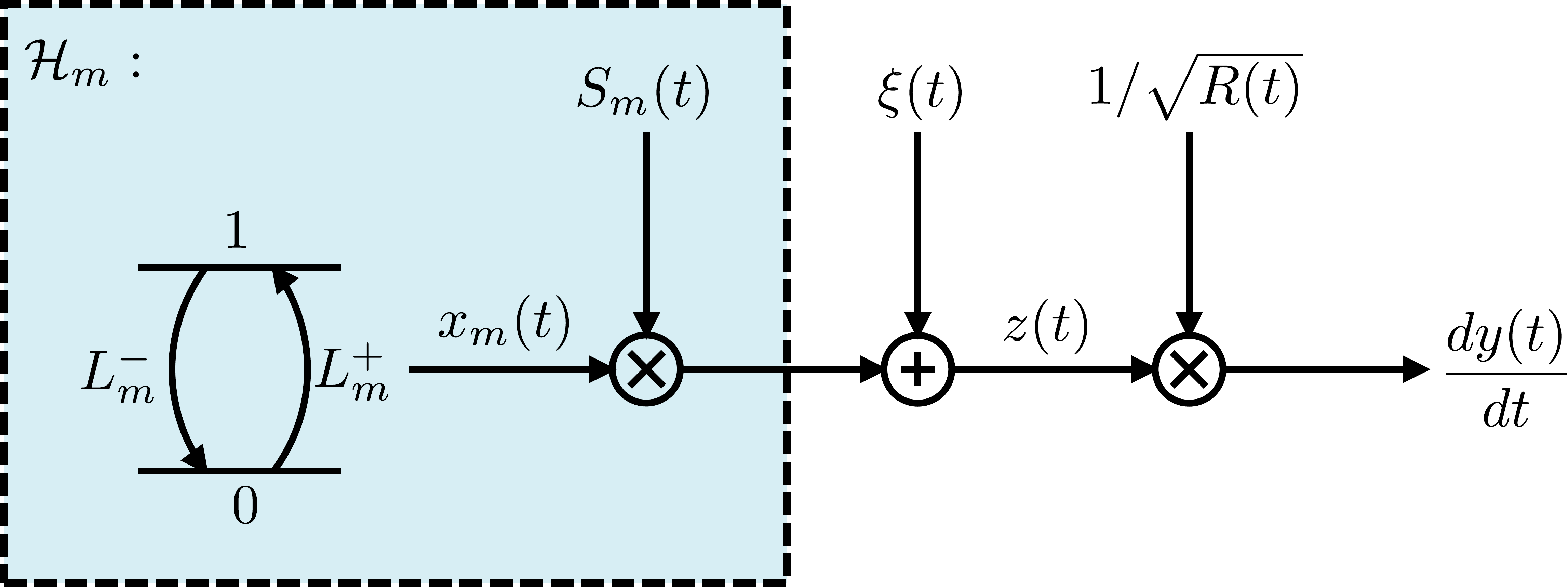}}
\caption{(Color online) A block diagram for the observation
  model. $\mathcal H_m$ is a hypothesis, $x_m(t)$ is the hidden
  signal, assumed here to be a two-state Markov process with
  transition rates $L_m^-$ and $L_m^+$, $S_m(t)$ is the signal
  amplitude, $\xi(t)$ is an additive white Gaussian noise, and $z(t)$
  is the observation process. The definition of observation processes
  $dy(t)/dt$ and $y(t)$, normalized with respect to the noise power
  $R(t)$, is for mathematical convenience.}
\label{observation}
\end{figure}

In stochastic detection theory, it is convenient to define a
normalized observation process $y(t)$ as the time integral of $z(t)$:
\begin{align}
y(t) &\equiv \int_0^t d\tau \frac{z(\tau)}{\sqrt{R(\tau)}},
\label{y}
\end{align}
and represent it using a stochastic differential equation:
\begin{align}
\mathcal H_m:\ dy(t) &\equiv y(t+dt) - y(t) 
\nonumber\\
&= dt \sigma_m(t)x_m(t) + dW(t),
\label{sde_y}
\\
\sigma_m(t) &\equiv \frac{S_m(t)}{\sqrt{R(t)}},
\label{sigma}
\end{align}
where $W(t)$ is the standard Wiener process with increment variance
$dW^2(t) = dt$ and It\={o} calculus \cite{gardiner,mao} is assumed
throughout this paper. The null hypothesis, in particular, is taken to
be
\begin{align}
\mathcal H_0: dy(t) &= dW(t).
\end{align}
Fig.~\ref{observation} depicts the observation
model through a block diagram.  

\subsection{\label{sec_formula}Estimator-correlator formula}
\begin{figure}[htbp]
\centerline{\includegraphics[width=0.48\textwidth]{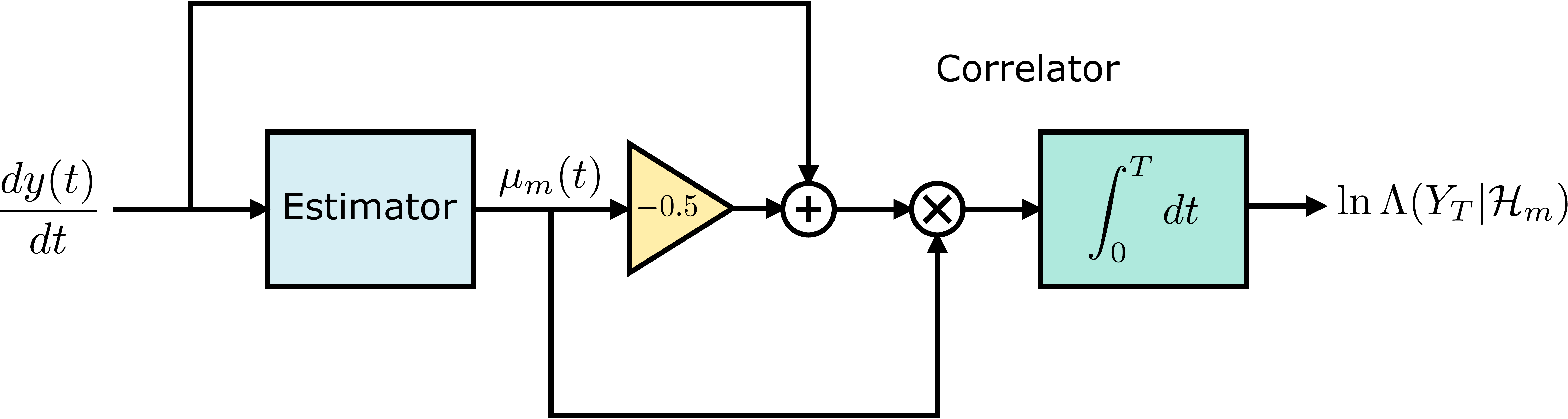}}
\caption{(Color online) An implementation of the estimator-correlator
  formula in Eq.~(\ref{formula}), which can be written as $\ln\Lambda
  = \int_0^T [dy(t)-dt\mu_m(t)/2]\mu_m(t)$.  $dy(t)$ in the integrand
  should be the future increment ahead of $t$ in accordance with
  It\={o} calculus.}
\label{estimator-correlator}
\end{figure}

Define the observation record as
\begin{align}
Y^T &\equiv \BK{y(t); 0\le t \le T}.
\label{Y}
\end{align}
Under rather general conditions about $x_m(t)$, the LLR $\ln
\Lambda(Y^T|\mathcal H_m)$ can be expressed using the
estimator-correlator formula
\cite{kailath_poor,duncan_lr,kailath69,hypothesis}, which correlates
the observation with an ``assumptive'' estimate $\mu_m(t)$:
\begin{align}
\ln \Lambda(Y^T|\mathcal H_m)
&= \int_0^T dy(t)\mu_m(t)
-\frac{1}{2}\int_0^T dt \mu_m^2(t),
\label{formula}
\end{align}
where 
\begin{align}
\mu_m(t) &\equiv \sigma_m(t)\expect\Bk{x_m(t)|Y^t,\mathcal H_m}
\label{assumptive}
\end{align}
is a causal estimator of the hidden signal conditioned on the
observation record $Y^t$ and the hypothesis $\mathcal H_m$. The
$dy(t)$ integral is an It\={o} integral, meaning that $dy(t)$ is the
future increment ahead of time $t$ and $\mu_m(t)$ in the integrand
$dy(t)\mu_m(t)$ should not depend on $dy(t)$. This rule is important
for consistent analytic and numerical calculations whenever one
multiplies $dy(t)$ with a signal that depends on $y(t)$
\cite{kailath_poor}.  Fig.~\ref{estimator-correlator} illustrates an
implementation of the formula.

As each $\ln\Lambda(Y^T|\mathcal H_m)$ depends only on one hypothesis
$\mathcal H_m$ (in addition to the fixed null hypothesis), once an
algorithm for its computation is implemented, it can be re-used even
if the other hypotheses are changed or new hypotheses are added. This
makes the estimator-correlator protocol more flexible and extensible
than the ones proposed in Refs.~\cite{gambetta07,danjou}, which are
specific to the hypotheses considered there.

Despite its simple appearance, the formula does not in general reduce
the complexity of the LLR calculation, as the estimator may still be
difficult to implement. We shall, however, present a simple numerical
method and some analytic solutions useful for the qubit readout
problem in the following.

\subsection{\label{qubit}Qubit dynamics}
For QND qubit readout, we assume that $x_m(t)$ is a classical
two-state first-order Markov process; Appendix~\ref{app_qnd} shows
explicitly how the classical theory can arise from the quantum
formalism of continuous QND measurement.  The possible values of
$x_m(t)$ are assumed to be
\begin{align}
x_m(t) &\in \BK{0,1}.
\end{align}
Other possibilities can be modeled by subtracting a baseline value
from the actual observation and defining an appropriate $\sigma_m(t)$
before the processing described here.  In the absence of measurements,
the probability function of $x_m(t) = x$ obeys a forward Kolmogorov equation
\cite{gardiner}:
\begin{align}
\frac{d\bs P_m(t)}{dt}
&= \bs L_m(t)\bs P_m(t),
\label{kolmogorov}\\
\bs P_m(t) &\equiv 
\bk{\begin{array}{c}P(x=0,t|\mathcal H_m)\\ 
P(x=1,t|\mathcal H_m)\end{array}},
\label{uncond_prob}\\
\bs L_m(t) &\equiv
\bk{\begin{array}{cc}-L_m^{+}(t) & L_m^{-}(t) \\ 
L_m^{+}(t) & -L_m^{-}(t)\end{array}},
\label{jump}
\end{align}
where $L_m^-$ and $L_m^+$ are the spontaneous decay and excitation
rates conditioned on the hypothesis and can be time-varying for
generality.  The decay time constant $1/L_m^-$ is commonly called
$T_1$, and $L_m^+$ can be used to model a random turn-on time
\cite{danjou}. For example, we can model the problem studied by
Gambetta and coworkers \cite{gambetta07} by defining
\begin{itemize}
\item $\mathcal H_0$: the qubit is in the $x=0$ state, and $x_0(t) = 0$.
\item $\mathcal H_1$: the qubit is in the $x=1$ state initially,
  $P(x=1,t=0|\mathcal H_1) = 1$, and the unconditional statistics of
  $x_1(t)$ obey Eqs.~(\ref{kolmogorov})--(\ref{jump}), with $L_1^-$
  being the decay rate and $L_1^+ = 0$.
\end{itemize}

\subsection{\label{estimator}Estimator}
\begin{figure}[htbp]
\centerline{\includegraphics[width=0.48\textwidth]{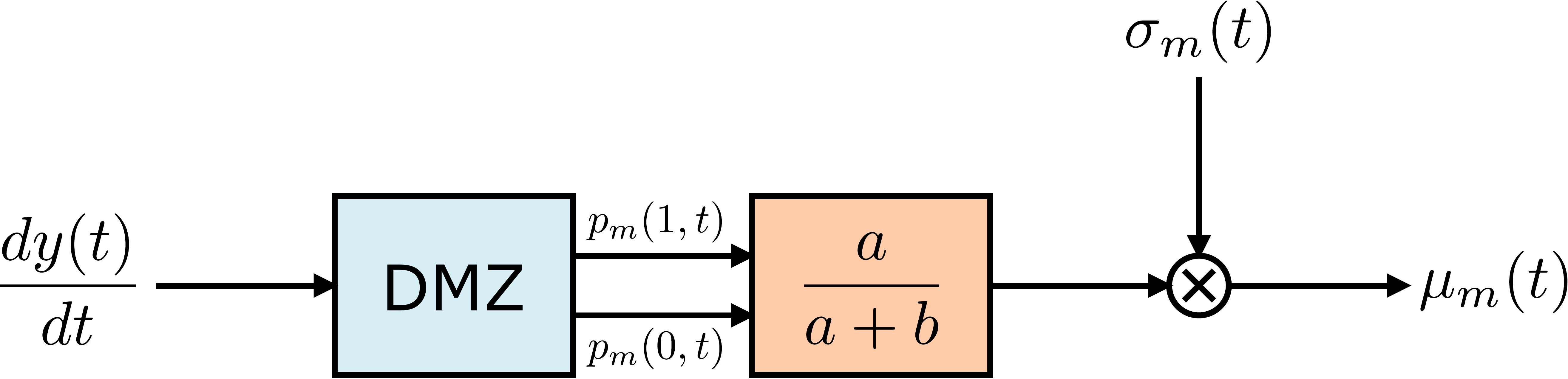}}
\caption{(Color online) A block diagram for the estimator using the
  Duncan-Mortensen-Zakai (DMZ) equation.}
\label{fig_dmz}
\end{figure}

The estimator $\mu_m(t)$ can be computed using the
Duncan-Mortensen-Zakai (DMZ) equation
\cite{duncan,mortensen,zakai,wong_hajek}:
\begin{align}
d\bs p_m(t) &= dt \bs L_m(t)\bs p_m(t) + dy(t) 
\sigma_m(t)\bs x\bs p_m(t),
\label{dmz}
\\
\bs p_m(t) &\equiv \bk{\begin{array}{c}p_{m}(x=0,t)\\ p_m(x=1,t)\end{array}},
\quad
\bs x \equiv \bk{\begin{array}{cc}0 & 0\\ 0 & 1\end{array}},
\label{pm}
\end{align}
where 
\begin{align}
p_m(x,t) \propto P(x,t|Y^t,\mathcal H_m)
\label{unnorm}
\end{align}
is the unnormalized posterior probability function of $x_m(t)$
conditioned on $Y^t$ and $\mathcal H_m$, and the initial condition is
determined by the initial prior probabilities:
\begin{align}
p_m(x,t=0) &=  P(x,t=0|\mathcal H_m).
\end{align}
The estimator is then
\begin{align}
\mu_m(t) &= \frac{\sigma_m(t) p_m(1,t)}{p_m(0,t)+p_m(1,t)},
\label{filter}
\end{align}
as depicted by Fig.~\ref{fig_dmz}. 

Although one can also use the Wonham equation \cite{wonham} to
perform the estimator, and the normalization step would not be
needed in theory, the DMZ equation is linear with respect to $\bs
p_m(t)$ and easier to solve analytically or numerically. In general, a
numerical split-step method can be used \cite{higham}:
\begin{align}
\bs p_m(t+dt) 
&\approx 
\exp \Bk{dy(t)\sigma_m(t)\bs x-\frac{dt}{2}\sigma_m^2(t)\bs x^2}
\nonumber\\&\quad\times
\exp\Bk{dt\bs L_m(t)}\bs p_m(t).
\end{align}
Many other numerical methods are available \cite{kloeden}.  Analytic
solutions can be obtained in the following special cases.

\subsection{\label{deterministic}Deterministic-signal detection}
For a simple example, assume binary hypothesis testing ($M = 2$), no
spontaneous transition ($L_m^- = L_m^+ = 0$), and deterministic
initial conditions given by
\begin{align}
p_0(0,0) = P(x=0,t=0|\mathcal H_0) &= 1,
\label{p0null}\\
p_0(1,0) = P(x=1,t=0|\mathcal H_0) &= 0,
\label{p1null}\\
p_1(0,0) = P(x=0,t=0|\mathcal H_1) &= 0,
\label{p0init}\\
p_1(1,0) = P(x=1,t=0|\mathcal H_1) &= 1.
\label{p1init}
\end{align}
The estimator becomes independent of the observation:
\begin{align}
\mu_0(t) &= 0,
&
\mu_1(t) &= \sigma_1(t).
\end{align}
This is simply a case of deterministic-signal detection, when the
estimator-correlator formula in Eq.~(\ref{formula}) becomes a matched
filter \cite{vantrees,kailath_poor}.  The minimum error probability
$P_{e,\textrm{min}}$ has an analytic expression \cite{vantrees}:
\begin{align}
P_{e,\textrm{min}} &= P_+ P(\mathcal H_0) + P_- P(\mathcal H_1),
\\
P_\pm &\equiv \frac{1}{2}\erfc\Bk{\sqrt{\frac{\textrm{SNR}}{8}}
\bk{1\pm\frac{2\lambda}{\textrm{SNR}}}},
\\
\erfc u &\equiv \frac{2}{\sqrt{\pi}}\int_u^\infty dv \exp(-v^2),
\\
\textrm{SNR} &\equiv \int_0^T dt \sigma_1^2(t),
\quad
\lambda \equiv \ln \frac{P(\mathcal H_1)}{P(\mathcal H_0)}.
\end{align}
For $\textrm{SNR}\to\infty$, the error exponent has the asymptotic
behavior $-\ln P_{e,\textrm{min}}\to \textrm{SNR}/8$.

Although this solution for $P_{e,\textrm{min}}$ is not strictly valid
when spontaneous transitions are present, it should be accurate when
the observation time $T$ is short relative to $1/L_1^-$ or $1/L_1^+$
and can serve as a rough guide for other cases.

\subsection{\label{noex}No spontaneous excitation ($L_m^+ = 0$)}
\begin{figure}[htbp]
\centerline{\includegraphics[width=0.48\textwidth]{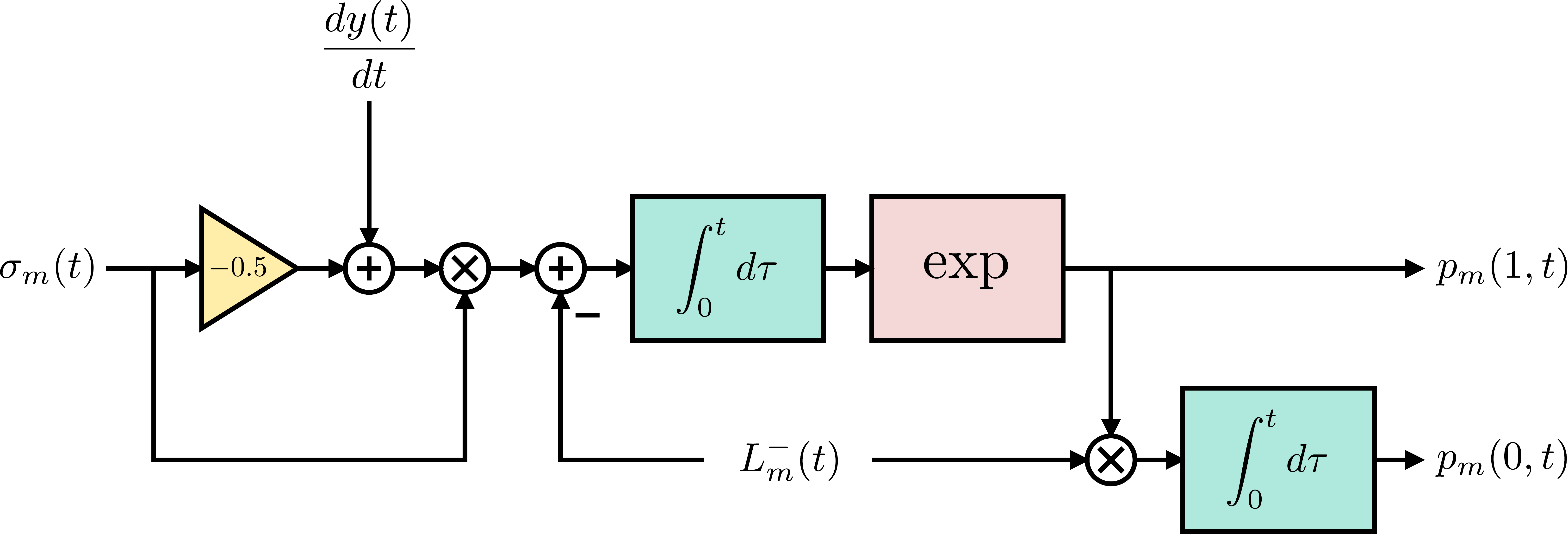}}
\caption{(Color online) Solution to the DMZ equation with spontaneous
  decay ($L_m^- > 0$), no spontaneous excitation ($L_m^+ = 0$), and an
  initial excited state ($p_m(1,t=0) = 1$, $p_m(0,t=0) = 0$).}
\label{fig_gbm}
\end{figure}

The case of $L_m^- > 0$ and $L_m^+ = 0$ corresponds to the model
studied by Gambetta and coworkers \cite{gambetta07}.  Eq.~(\ref{dmz})
becomes
\begin{align}
dp_m(0,t) &=  dt L_m^-(t) p_m(1,t),
\label{gbm0}\\
dp_m(1,t) &= -dt L_m^-(t) p_m(1,t) + dy(t)\sigma_m(t)p_m(1,t).
\label{gbm}
\end{align}
Eq.~(\ref{gbm}) describes the famous geometric Brownian motion
\cite{mao}. Its well known solution can be obtained by applying
It\={o}'s lemma to $d\ln p_m(1,t)$ and is given by
\begin{align}
p_m(1,t) &= p_m(1,0)
\exp\left\{\int_0^t dy(\tau)\sigma_m(\tau)\right.
\nonumber\\&\quad
\left.-\int_0^t d\tau\Bk{\frac{\sigma_m^2(\tau)}{2}+L_m^-(\tau)}
\right\}.
\label{fm1}
\end{align}
A time integral of $p_m(1,t)$ then gives $p_m(0,t)$:
\begin{align}
p_m(0,t) &= p_m(0,0) + \int_0^t d\tau L_m^-(\tau)p_m(1,\tau).
\label{fm0}
\end{align}
For binary qubit state discrimination, we can assume that $\mu_0(t) =
0$, and $\mu_1(t)$ can be determined from Eqs.~(\ref{fm1}),
(\ref{fm0}), and (\ref{filter}), starting from the deterministic
initial conditions given by Eqs.~(\ref{p0init}) and (\ref{p1init}) if
the measurement starts immediately after the qubit state is prepared,
as shown in Fig.~\ref{fig_gbm}. If there is a finite arming time
before the measurement starts \cite{gambetta07,danjou}, the forward
Kolmogorov equation (\ref{kolmogorov}) can be used to determine the
initial state probabilities.

\subsection{\label{nodecay}No spontaneous decay ($L_m^- = 0$)}
One can assume $L_m^+>0$ and $L_m^-=0$ to model a random signal
turn-on time \cite{danjou} and negligible spontaneous decay ($T \ll
1/L_m^-$). The simplest way of computing $\mu_m(t)$ is to define a new
observation process
\begin{align}
dy'(t) &\equiv dy(t)-\sigma_m(t)dt
\nonumber\\
&=-dt\sigma_m(t)\Bk{1-x_m(t)} + dW(t).
\label{dyp}
\end{align}
A new DMZ equation can then be expressed in terms of $y'(t)$ and is
given by
\begin{align}
dp_m(0,t) &= -dt L_m^+(t) p_m(0,t)-dy'(t)\sigma_m(t)p_m(0,t),
\\
dp_m(1,t) &= dt L_m^+(t)p_m(0,t),
\end{align}
which have the same form as Eqs.~(\ref{gbm0}) and (\ref{gbm}) and can
be solved using the same method. The final solution is
\begin{align}
p_m(0,t) &= p_m(0,0)\exp\left\{-\int_0^t dy(\tau)\sigma_m(\tau)
\right.
\nonumber\\&\quad
\left.+\int_0^t d\tau \Bk{\frac{\sigma_m^2(\tau)}{2}-L_m^+(\tau)}\right\},
\\
p_m(1,t) &= p_m(1,0) + \int_0^td\tau L_m^+(\tau)p_m(0,\tau).
\end{align}

\section{\label{poisson}Poissonian noise model}
\subsection{\label{process_poisson}Observation process}
For photon-counting measurements, it is more appropriate to assume
that the counting process $n(t)\in\{0,1,2,\dots\}$,
conditioned on the hidden process $X_m^t\equiv\{x_m(\tau);0\le\tau\le
t\}$, obeys Poissonian statistics \cite{snyder_miller}:
\begin{align}
&\quad P(n(t)|X_m^t,\mathcal H_m) 
\nonumber\\
&= \exp\Bk{-\int_0^t d\tau\lambda_m(\tau)}
\frac{\Bk{\int_0^t d\tau \lambda_m(\tau)}^{n(t)}}{n(t)!},
\label{poisson_dist}
\end{align}
where 
\begin{align}
\lambda_m(t) &\equiv \lambda_0(t)\Bk{1+\alpha_m(t)x_m(t)}
\label{alpha}
\end{align} 
is the intensity of the Poisson process and $\alpha_m(t)$ is a
deterministic signal amplitude. $dn(t)\in\{0,1\}$ is then the detected
photon number at time $t$. We assume $\mathcal H_0$ with known
intensity $\lambda_0(t)>0$ to be the null hypothesis.


\begin{figure}[htbp]
\centerline{\includegraphics[width=0.48\textwidth]{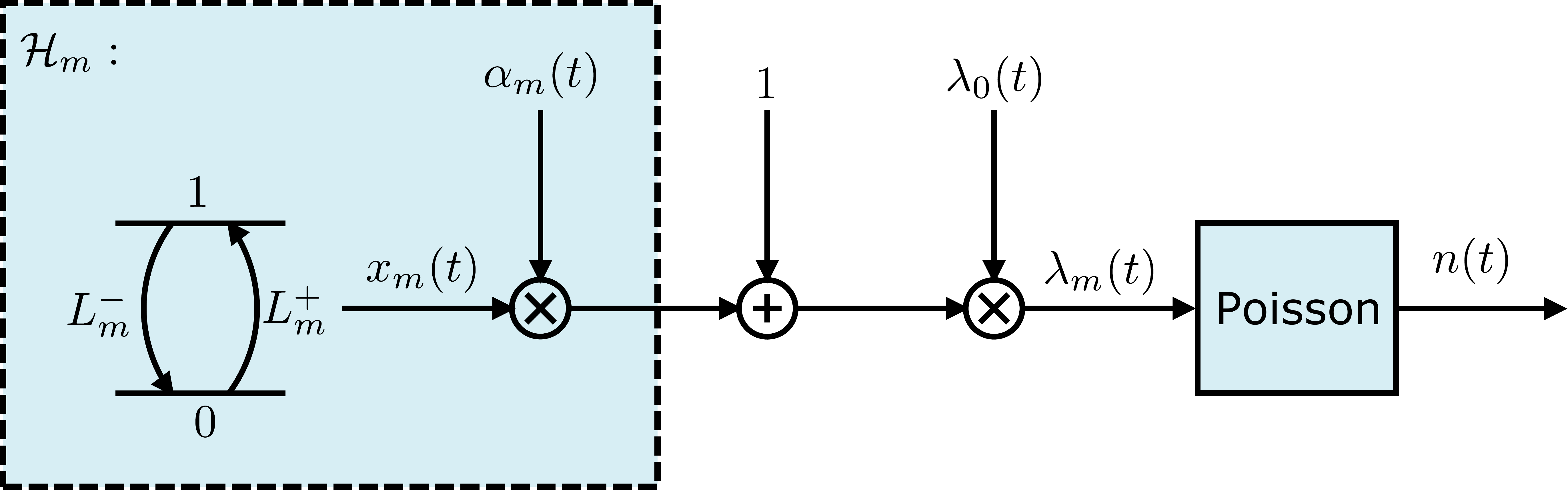}}
\caption{(Color online) The Poissonian observation model. The counting
  process $n(t)$ is driven by the stochastic intensity
  $\lambda_m(t)$.}
\label{observation_poisson}
\end{figure}
\subsection{\label{sec_formula_poisson}Estimator-correlator formula}
\begin{figure}[htbp]
\centerline{\includegraphics[width=0.48\textwidth]{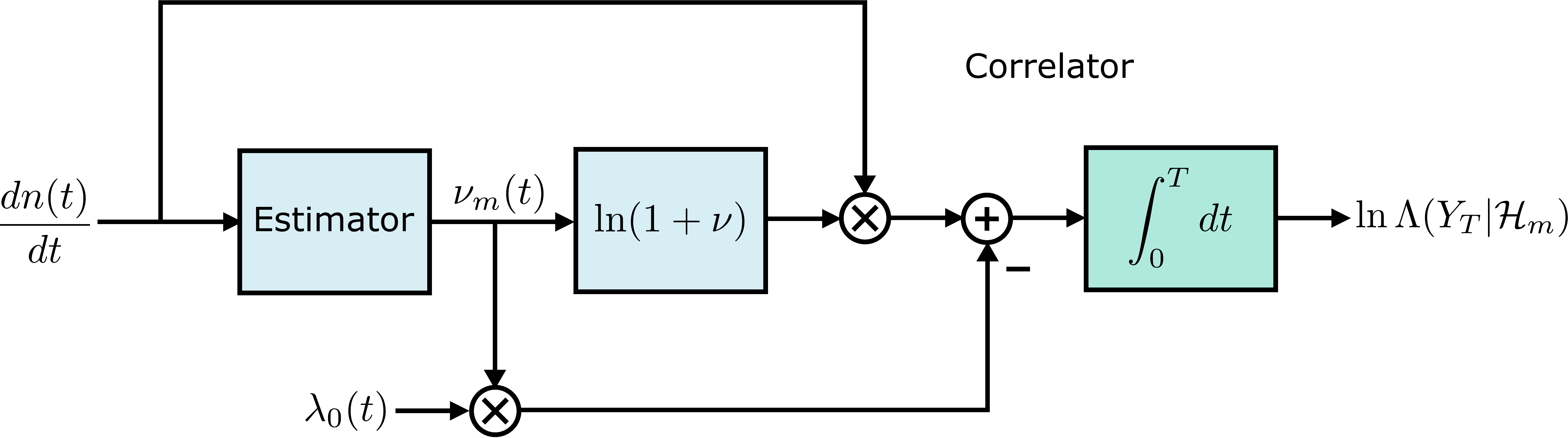}}
\caption{(Color online) The estimator-correlator structure for the
  Poissonian observation model. $dn(t)$ should be the future increment
  ahead of $t$ when multiplied with $\ln[1+\nu_m(t)]$.}
\label{fig_formula_poisson}
\end{figure}

Define the observation record as
\begin{align}
N^T &\equiv \BK{n(t); t_0 \le t \le T}.
\label{N}
\end{align}
Our goal is to calculate the LLR
\begin{align}
\ln\Lambda(N^T|\mathcal H_m) &= \ln \frac{dP(N^T|\mathcal H_m)}{dP(N^T|\mathcal H_0)}.
\end{align}
A formula analogous to the Gaussian case in Eq.~(\ref{formula}) is
given by \cite{snyder,hypothesis}
\begin{align}
\ln\Lambda(N^T|\mathcal H_m)
&= \int_0^T dn(t) \ln \Bk{1+\nu_m(t)}
\nonumber\\&\quad
-\int_0^T dt \lambda_0(t)\nu_m(t),
\label{formula_poisson}
\\
\nu_m(t) &\equiv \alpha_m(t)\expect\Bk{x_m(t)|N^t,\mathcal H_m},
\label{nu}
\end{align}
where the $dn(t)$ integral should again follow It\={o}'s convention
\cite{snyder_miller}. Fig.~\ref{fig_formula_poisson} illustrates the formula.

\subsection{\label{estimator_poisson}Estimator}
We assume the same unconditional qubit dynamics described in
Sec.~\ref{qubit}.  The estimator can be computed from a DMZ-type
equation \cite{wong_hajek,hypothesis}:
\begin{align}
d\bs p_m(t) &= dt \bs L_m(t)\bs p_m(t) 
+ \Bk{dn(t)-dt\kappa(t)}
\nonumber\\&\quad\times
\BK{\frac{\lambda_0(t)}{\kappa(t)}
\Bk{\bs I+\alpha_m(t)\bs x}-\bs I}\bs p_m(t),
\label{dmzp}
\end{align}
where $\kappa(t) > 0$ is an arbitrary positive reference intensity and
the estimator is
\begin{align}
\nu_m(t) &= \frac{\alpha_m(t)p_m(1,t)}{p_m(0,t)+p_m(1,t)}.
\end{align}
This procedure is identical to that depicted in Fig.~\ref{fig_dmz}.
Assuming $\kappa(t) = \lambda_0(t)$, Eq.~(\ref{dmzp}) can be solved
using a numerical split-step method:
\begin{align}
&\quad
\bs p_m(t+dt)
\nonumber\\ 
&\approx \exp\BK{dn(t)\ln\Bk{\bs I+\alpha_m(t)\bs x}-dt\lambda_0(t)\alpha_m(t)\bs x}
\nonumber\\&\quad\times
\exp\Bk{dt\bs L_m(t)}\bs p_m(t).
\end{align}
Analytic solutions can be found in the following cases.

\subsection{\label{noex_poisson}No spontaneous excitation ($L_m^+ = 0$)}
\begin{figure}[htbp]
\centerline{\includegraphics[width=0.48\textwidth]{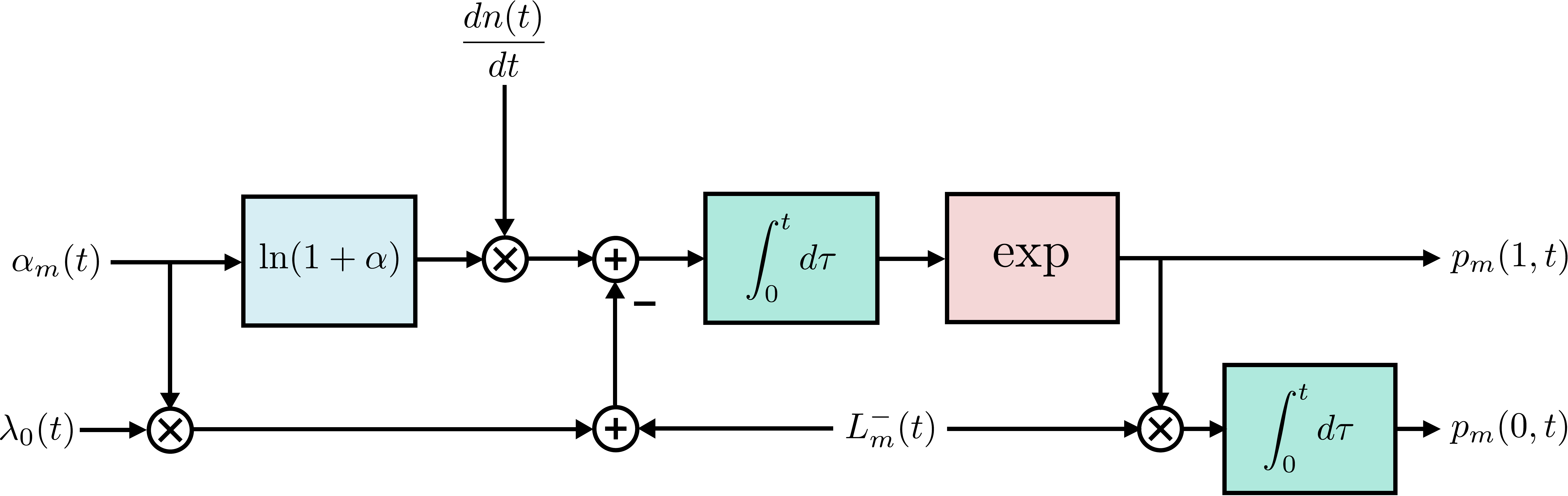}}
\caption{(Color online) A block diagram for Eqs.~(\ref{pm1sol}) and
  (\ref{pm0sol}), a solution to the Poissonian DMZ equation.  $L_m^+ =
  0$, $p_m(1,t=0) = 1$, and $p_m(0,t=0) = 0$ are assumed.}
\label{dmz_poisson}
\end{figure}

Let $\kappa(t) = \lambda_0(t)$. Eq.~(\ref{dmzp}) becomes
\begin{align}
dp_m(0,t) &= dt L_m^-(t) p_m(1,t),
\label{dmzp0}\\
dp_m(1,t) &= -dt L_m^-(t)p_m(1,t)
\nonumber\\&\quad
+ \Bk{dn(t)-dt\lambda_0(t)}\alpha_m(t)p_m(1,t).
\label{dmzp1}
\end{align}
Following Chap.~5.3.1 in Ref.~\cite{snyder_miller}, we get
\begin{align}
p_m(1,t) &= p_m(1,0)\exp\left\{\int_0^t dn(\tau)
\ln\Bk{1+ \alpha_m(\tau)}\right.
\nonumber\\&\quad
\left.
-\int_0^t d\tau\Bk{\lambda_0(\tau)\alpha_m(\tau)+L_m^-(\tau)}\right\},
\label{pm1sol}\\
p_m(0,t) &= p_m(0,0)+\int_0^t d\tau L_m^-(\tau)p_m(1,\tau).
\label{pm0sol}
\end{align}
Fig.~\ref{dmz_poisson} depicts a block diagram for this solution.

\subsection{\label{nodecay_poisson}No spontaneous decay ($L_m^- = 0$)}
We now let $\kappa(t) = \lambda_0(t)\Bk{1+\alpha_m(t)}$.
Eq.~(\ref{dmzp}) becomes
\begin{align}
dp_m(0,t) &= -dt L_m^+(t)p_m(0,t)
\nonumber\\&\quad-
\Bk{dn(t)-dt\kappa(t)}
\frac{\alpha_m(t)}{1+\alpha_m(t)} p_m(0,t),
\\
dp_m(1,t) &= dtL_m^+(t)p_m(0,t).
\end{align}
Similar to the previous case, the solution is
\begin{align}
p_m(0,t) &= p_m(0,0)\exp\left\{-\int_0^t dn(\tau)\ln\Bk{1+\alpha_m(\tau)}
\right.
\nonumber\\&\quad
+\left.\int_0^t d\tau\Bk{\lambda_0(\tau)\alpha_m(\tau)-L_m^+(\tau)}\right\},
\\
p_m(1,t) &= p_m(1,0) + \int_0^t d\tau L_m^+(\tau)p_m(0,\tau).
\end{align}

It is interesting to note that all the Poissonian results approach the
Gaussian ones in Sec.~\ref{gauss} if we assume $dn =
\sqrt{\lambda_0}dy+ \lambda_0 dt$, $\alpha_m
=\sigma_m/\sqrt{\lambda_0}$, and $\lambda_0\to\infty$.

\section{Conclusion}
We have proposed an estimator-correlator architecture for optimal
qubit-readout signal processing and found analytic solutions in some
special cases of interest using It\={o} calculus.  Although we have
focused on a classical model, our formalism can potentially be
extended to more general quantum dynamics \cite{hypothesis,cook14} and
more realistic measurements, including artifacts such as dark counts
and finite detector bandwidth \cite{wiseman_milburn}.  An open problem
of interest is the evaluation of readout performance beyond the case
of deterministic-signal detection. Numerical Monte Carlo simulation is
not difficult for two-level systems, but analytic solutions should
bring additional insight and may be possible using tools in classical
and quantum detection theory
\cite{vantrees,bucklew,helstrom,tsang_nair,mismatch,testing_quantum}.
Another open problem is the accuracy, speed, and practicality of our
algorithms in reality, which will be subject to more specific
experimental requirements and hardware limitations \cite{stockton02}.

\section*{Acknowledgments}
This work is supported by the Singapore National Research Foundation
under NRF Grant No.~NRF-NRFF2011-07.

\appendix
\section{\label{app_qnd}Quantum formalism of
continuous quantum-nondemolition measurement}
Let
\begin{align}
\hat f_m(t) &= \bk{\begin{array}{cc}f_m(0,0,t) & f_m(0,1,t)\\
f_m(1,0,t) & f_m(1,1,t)\end{array}}
\end{align}
be the unnormalized density matrix for the qubit conditioned on the
observation record $Y^t$ and hypothesis $\mathcal H_m$.  Consider the
following linear stochastic quantum master equation
\cite{wiseman_milburn}:
\begin{align}
d\hat f_m &= dt L_m^-\bk{\hat \sigma_- \hat f_m\hat \sigma_+
-\frac{1}{2}\hat \sigma_+\hat \sigma_- \hat f_m
-\frac{1}{2}\hat f_m\hat \sigma_+\hat \sigma_-}
\nonumber\\&\quad
+dt L_m^+\bk{\hat \sigma_+\hat f_m\hat \sigma_--\frac{1}{2}\hat\sigma_-\hat\sigma_+ \hat f_m
-\frac{1}{2}\hat f_m\hat\sigma_-\hat\sigma_+}
\nonumber\\&\quad
+dt L_m^x \bk{\hat x \hat f_m \hat x-\frac{1}{2}\hat x^2 \hat f_m 
- \frac{1}{2}\hat f_m\hat x^2}
\nonumber\\&\quad
+ \frac{dy \sigma_m}{2}\bk{\hat x \hat f_m + \hat f_m \hat x},
\label{qsme}
\end{align}
where
\begin{align}
\hat\sigma_- &= \bk{\begin{array}{cc}0&1\\0&0\end{array}},
&
\hat\sigma_+ &= \bk{\begin{array}{cc}0&0\\1&0\end{array}},
&
\hat x &= \bk{\begin{array}{cc}0 & 0\\0& 1\end{array}},
\end{align}
and $L_m^-$, $L_m^+$, and $L_m^x \ge \sigma_m^2/4$ are the decay,
excitation, and dephasing rates, respectively.  The estimator in the
quantum estimator-correlator formula \cite{hypothesis} is
\begin{align}
\sigma_m(t)\expect\bk{\hat x|Y^t,\mathcal H_m} &= 
\frac{\sigma_m(t)f_m(1,1,t)}{f_m(0,0,t)+f_m(1,1,t)}.
\label{qest}
\end{align}
The important point here is that the estimator involves only the
diagonal components of $\hat f_m(t)$, which are decoupled from the
off-diagonal components throughout the evolution:
\begin{align}
df_m(0,0,t) &= dt\Bk{-L_m^+(t) f_m(0,0,t)+L_m^-(t) f_m(1,1,t)},
\label{fm00}\\
df_m(1,1,t) &= dt\Bk{L_m^+(t) f_m(0,0,t)-L_m^-(t) f_m(1,1,t)}
\nonumber\\&\quad
+dy(t)\sigma_m(t)f_m(1,1,t).
\label{fm11}
\end{align}
This means that a classical stochastic model is sufficient. In
particular, Eqs.~(\ref{fm00}) and (\ref{fm11}) are identical to the
classical DMZ equation given by Eq.~(\ref{dmz}).  The argument in the
case of Poissonian noise is similar.  \bibliography{research}

\end{document}